\documentclass[a4paper]{spie}  

\usepackage{amsmath,amsfonts,amssymb}
\usepackage{graphicx}
\usepackage[colorlinks=true, allcolors=blue]{hyperref}

\title{Astronomical interferometry with near-IR e-APD at CHARA: characterization, optimization and on-sky operation}

\author[a]{Cyprien Lanthermann}
\author[a,b]{Jean-Baptiste Le Bouquin}
\author[c]{Narsireddy Anugu}
\author[b]{John Monnier}
\author[c]{Stefan Kraus}
\affil[a]{Institut de Planetologie et d'Astrophysique de Grenoble , 414 Rue de la Piscine, 38400 Saint-Martin-d'Hères, France}
\affil[b]{University of Michigan, Department of Astronomy, MI, United States}
\affil[c]{University of Exeter, School of Physics and Astronomy, Stocker Road, Exeter, EX4 4QL}

\begin{document}

\maketitle

\begin{abstract}
We characterize a near-infrared C-RED ONE camera from First Light Imaging (FLI). This camera uses a SAPHIRA electron avalanche photo-diode array (e-APD) from Leonardo (previously Selex). To do so, we developed a model of the signal distribution. This model allows a measurement of the gain and the Excess Noise Factor (ENF) independently of preexisting calibration such as the system gain. The results of this study show a gain which is 0.53 $\pm$ 0.04 times the gain reported by the manufacturer. The measured ENF is 1.47$\pm$ 0.03 when we expected 1.25. For an avalanche gain of $\simeq$ 60 and a frame rate larger than 100 Hz, the total noise can be lower than 1 $e^{-}$/frame/pixel. The lowest dark current level is 90$e^{-}$/s/pixel, in agreement with the expected H-band background passing through the camera window. These performance values provide a significant improvement compared to earlier-generation PICNIC camera and allowed us to improve the performance of the Michigan infrared combiner (MIRC) instrument at the Center for High Angular Resolution Astronomy (CHARA), as part of our MIRC-X instrumentation project.
\end{abstract}


\section{INTRODUCTION}\label{sec:intro}
Optical and infrared interferometry is an observation technique that permits to study astronomical bodies with a better angular resolution than with direct imagery observation at the same wavelength. But the interferometry suffers from a sensitivity limit.
In the near-infrared, the typical coherence time is of few milliseconds. It corresponds to a sky background of $<$ 0.1 ph per exposure for a diffraction-limited instrument operating in H-band (1.65 $\mu m$). With $\simeq$ 10 $e^{-}$ readout noise at these frame rates, classical near-infrared cameras are the limiting factor for sensitivity.

To improve the sensitivity of near-infrared interferometers, we need fast (frame rate $>$ 200Hz), sensitive (quantum efficiency $>$ 50$\%$), large format ($>$ 200 x 200 pixels) and low readout noise ($<$ 1$e^{-}$/frame/pixel) detectors. For this qualities, the electron avalanche photodiode arrays (e-APD\cite{gert2010, gert2012}) present a breakthrough.
That is why the e-APD technology has been chosen to improve the sensitivity of the Center for High Angular Resolution Astronomy (CHARA~\cite{CHARA}) array (Mount Willson Observatory, USA) with the implementation of two C-RED ONE~\cite{CRED} cameras, from First Light Imaging (FLI). One camera is used for operations in the H-band in the instrument MIRC-X\cite{MIRCX}, the successor of the Michigan infrared combiner (MIRC~\cite{MIRC}), and the other is used for operations in the K-band in the upcoming instrument MYSTIC\cite{MYSTIC}. The H-band camera for MIRC-X sees a warm (300K) spectrograph at f/4, while the K-band camera for MYSTIC will be implemented in a 220K cryostat to minimize the instrument emission. The results in this paper are focused on the H-band camera of MIRC-X.

\section{e-APD technology}\label{sec:apdtech}

The e-APD~\cite{gert2010, gert2012} technology applies a bias voltage in a lower layer of the pixel called multiplication region. This bias voltage accelerates the electron, generated by a photon in the upper layer of the detector. This electron migrates to the bottom of the pixel through this multiplication region. The energy obtained by the acceleration becomes sufficient to ionize an atom of the substrate by collision, generating a second electron. The first and the second electrons are accelerated again, generating others electrons and creating a chain reaction. 

This process multiplies the photonic signal directly in the pixel, therefore increasing the signal to noise ratio (SNR) with respect to the readout noise. With a multiplexing readout system, we can read simultaneously several pixels, and hence increase the overall maximum frame rate.
As the avalanche gain results in a process of impact ionization in the substrate of the detector, the total gain presents a certain probability distribution (M). The Excess Noise Factor (ENF) is then an additional noise and a consequence of this distribution. In the literature \cite{n-photon, 4711091}, the gain is defined by the mean value of the gain distribution 

\begin{equation}\label{eq:GM}
G = \langle M \rangle
\end{equation}
 and the ENF is defined by 
\begin{equation}\label{eq:ENFM}
ENF = \frac{\langle M ^{2}\rangle}{\langle M \rangle ^{2}}.
\end{equation} 

\subsection{C-RED ONE SPECIFICITIES}\label{sec:CREDspec}

The C-RED ONE is a fully-integrated camera system provided by FLI. It contains a SAPHIRA Mark13 model of an e-APD array made by Leonardo (previously Selex). This detector is placed in a vacuum container ($10^{-5}$ atm in operation) and is cooled by a pulsed tube (80 K in operation). The camera window is open at f/4, and the thermal emission at wavelengths larger than 1.9 $\mu$m is filtered by a set of four cold filters.
The characteristics of the camera given by FLI\cite{CRED-ENF} are summarised in Tab.~\ref{tab:FLIspec}.

\begin{table}[!h]
\caption{Characteristics of the C-RED ONE camera} 
\label{tab:FLIspec}
\begin{center}       
\begin{tabular}{|l|l|} 
\hline
\rule[-1ex]{0pt}{3.5ex}  Detector type & SAPHIRA MCT SWIR Mark13 – eAPD 320X256  \\
\hline
\rule[-1ex]{0pt}{3.5ex}  Wavelength ($\mu$m) & 0.8 to 2.5  \\
\hline
\rule[-1ex]{0pt}{3.5ex}  Read Out Speed, full frame single read (frame/s) & 3500  \\
\hline
\rule[-1ex]{0pt}{3.5ex} Total noise (1 ms of integration, 300 K scene) ($e^{-}/pixel$) & 0.5  \\
\hline
\rule[-1ex]{0pt}{3.5ex}  Quantum Efficiency ($\%$) & $>$ 70  \\
\hline 
\rule[-1ex]{0pt}{3.5ex} Excess Noise Factor (ENF) & 1.25 \\
\hline
\end{tabular}
\end{center}
\end{table}

The Mark13 model detector consists of ten blocks of 32 readout outputs, reading 32 x 256 (columns x rows) pixels each. These multiple outputs permit the multiplexing of the detector and to obtain a fast read capability, up to 3500~Hz for full-frame and 1 read per frame. The readout system allows reducing the readout window by choosing the output we want to read (columns) and the number of rows we want to read. The windowing increases the readout speed accordingly.

The C-RED ONE camera has several readout modes implemented, including a single read mode, a correlated double sampling (CDS) mode, a multiple non-destructive reads mode and a "IOTA" readout mode.
The single read mode reads the entire window once then reset it immediately. The integration time is the time between two frames. It is limited by the readout time of the window.
The CDS mode reads the entire window immediately after a frame demand then reset it and automatically read it again. The result of the CDS processing is the difference between those two frames. The maximum speed is twice lower than the maximum frame rate in single read mode.
The multiple non-destructive reads mode starts with a reset period and then a user-specified number of frames are read and delivered at maximum speed.
The IOTA mode is inspired by a readout scheme that was implemented at the PICNIC camera\cite{2004PASP..116..377P} of the IONIC3 instrument at the IOTA array. It consists of a multiple (Nreads) consecutive non-destructive reads of the same pixel. Then it reads the next pixel of the same row. It performs those reads a multiple (Nloops) time for the same line before reading the next line. At the end of a frame, the reads start over again with the first pixel of the window. The window is reset after that a multiple (Nframes) frames have been read. Nreads, Nloops, and Nframes are set by the user.

We personally requested to FLI to implement the IOTA mode to our cameras. It's the one we focus on for the on-sky operation.

\section{MODELING THE PHOTON DISTRIBUTION}\label{sec:model}

\subsection{Description of our Model}\label{sec:justimod}

It exists several ways to measure the characteristics of an e-APD detector. However, the classical methods rely on previous calibrations such as the system gain. In this section, we present our method to measure the total gain (including amplification) and the ENF, in a self-consistent way.

\subsubsection{Filtering the parasitic electronic signal}\label{sex:lookdata}

\begin{figure}[!h]
\centering
\includegraphics[width=\textwidth]{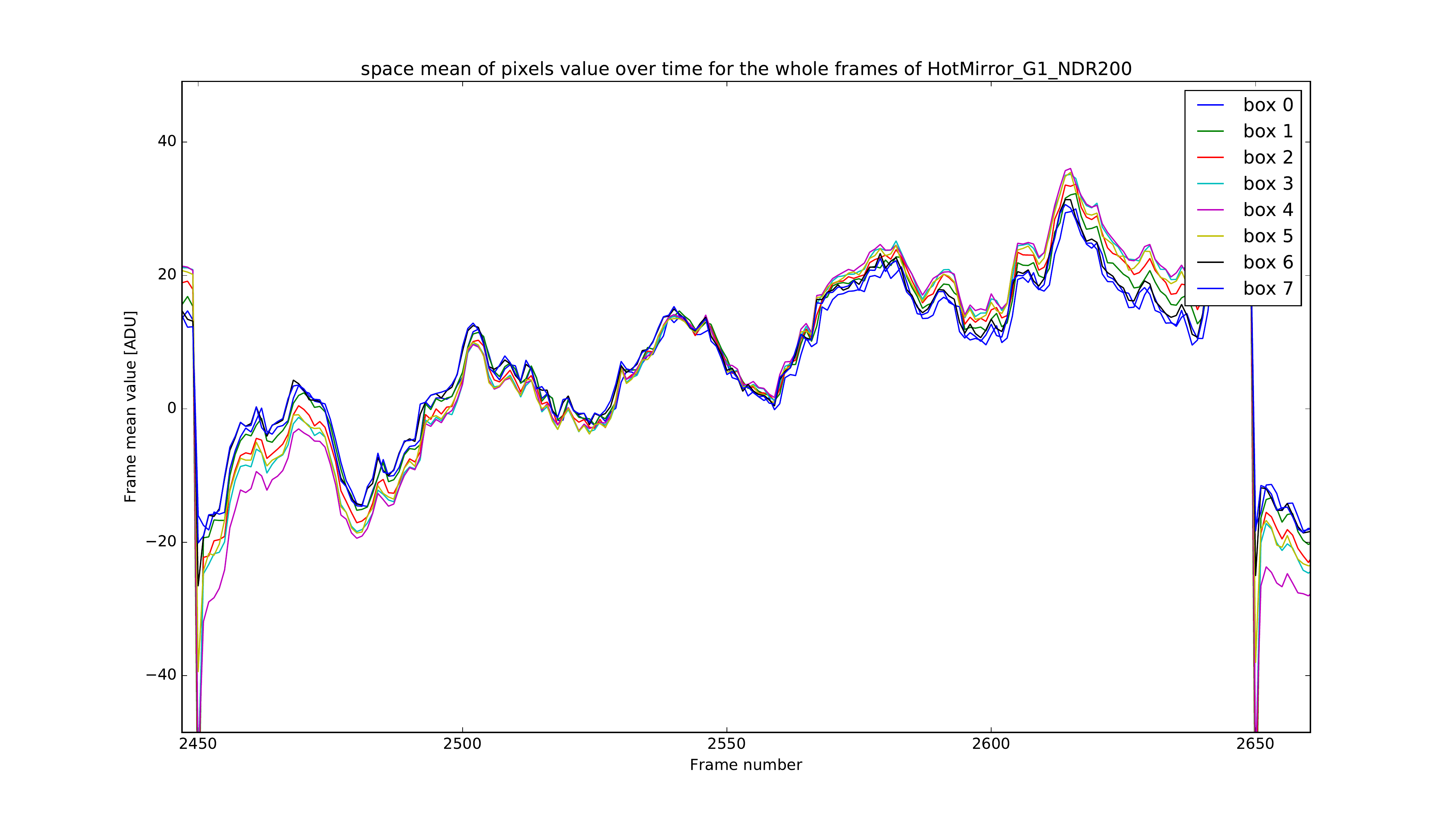}
\caption{Frame mean value versus frame number when observing a room temperature mirror with 200 nondestructive reads. The x-axis is zoomed between two resets.}\label{fig:ramp}
\end{figure}

In Fig.~\ref{fig:ramp}, the x-axis is the frame number and the y-axis is the mean value of a sub-window of 320~x~32~pixels, called box, and represented with different colors. The mean value of the ramp is subtracted to recenter the different boxes on the same level. The maximum frame rate is 3500 Hz. There are 200 non-destructive read between two resets. The detector is reset at the frames number 2450 and 2650. These data have been taken watching a mirror at room temperature, in order to minimize the amount of incident flux.

We see that the first frame after a reset (2450) presents an anomaly that we call the reset anomaly. This frame is discarded in our analysis and in the pipeline of the instrument. We also see that during the integration the mean value of a box presents a sort of sinusoidal signal combined with a linear trend. The linear trend is due to the accumulating dark current. The sinusoidal signal is a parasitic electronic signal. This electronic signal actually comes from the pulsed tube that cools the detector. This signal has a frequency of $\simeq$ 90 Hz and an amplitude of $\simeq$ 20 ADU. FLI didn't manage to get rid of it yet.

\subsubsection{Extracting temporal sequence}\label{sec:extracttempseq}

We collect data with the instrument configuration used for scientific operations but looking at an internal light. On that configuration, the image of a photometric channel consists in a band of $\simeq$ 3 pixels large and a dozen of pixels long. The dozen of pixels is actually the spectral dispersion of the light source. Because we are observing an internal light, the flux in the image is stable over time.

We use the IOTA readout mode, as for the sky observation. However, we slightly reduce the number of Nreads (8), Nloops (2) and the size of the window (320 x 20 pixels). These numbers were selected to achieve the highest possible maximum frame rate (1916 FPS), while still achieving a theoretical sub-electron readout noise. To our knowledge and understanding, this setup represents our best chance to measure individual photon in the camera signal. We use non-destructive reads of 50 frames between two resets. As said previously, we discard the first frame after reset.

We extract the flux of one pixel of a photometric channel by subtracting each frame value by its precedent. To get rid of the electronic sinusoidal signal (frequency $\simeq$ 90 Hz), this same operation is performed for several non-illuminated pixels in the same line of the extracted pixel. The median value of these pixels is subtracted from the value of the extracted pixel. This is possible because the maximum frame rate is high enough so we can consider that the sinusoidal signal is the same for a whole line (20 lines sample at $\simeq$ 1900 Hz, compared to a signal with a frequency of 90 Hz).
The final product is the temporal sequence of one pixel, in ADU.

We tried different technics to subtract the parasitic signal, by using the mean or the median of many reference non-illuminated pixels (typically more than 20). We also tried filtering the 90 Hz signal from the signal of the studied pixel itself. We verified that the final results of our analysis are unaffected by this choice.

\begin{figure}[!h]
\centering
\includegraphics[scale=0.65]{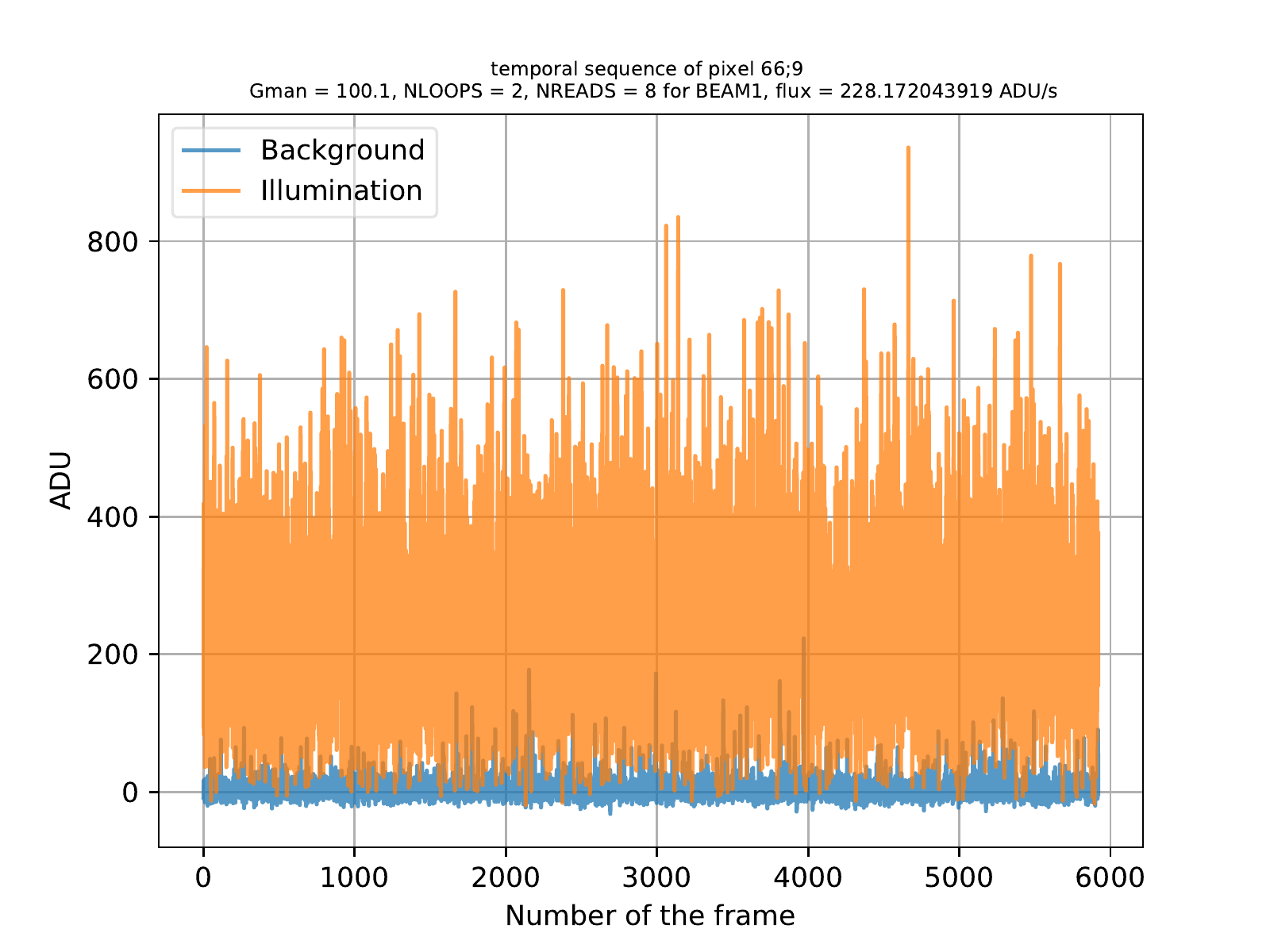}
\caption{Example of temporal sequence}\label{fig:dtempseq}
\end{figure}

Figure~\ref{fig:dtempseq} represents the pixel value in ADU versus the frame number. The temporal sequence when there is no illumination (Background) is in blue. The temporal sequence for the same pixel but with illumination is in orange. These additional photons are limited to the 1.6-1.65$\mu$m band-pass, thanks to the dispersive prism of the instrument.

\subsubsection{Histograms}\label{sec:histo}

\begin{figure}[!h]
\centering
\includegraphics[scale=0.6]{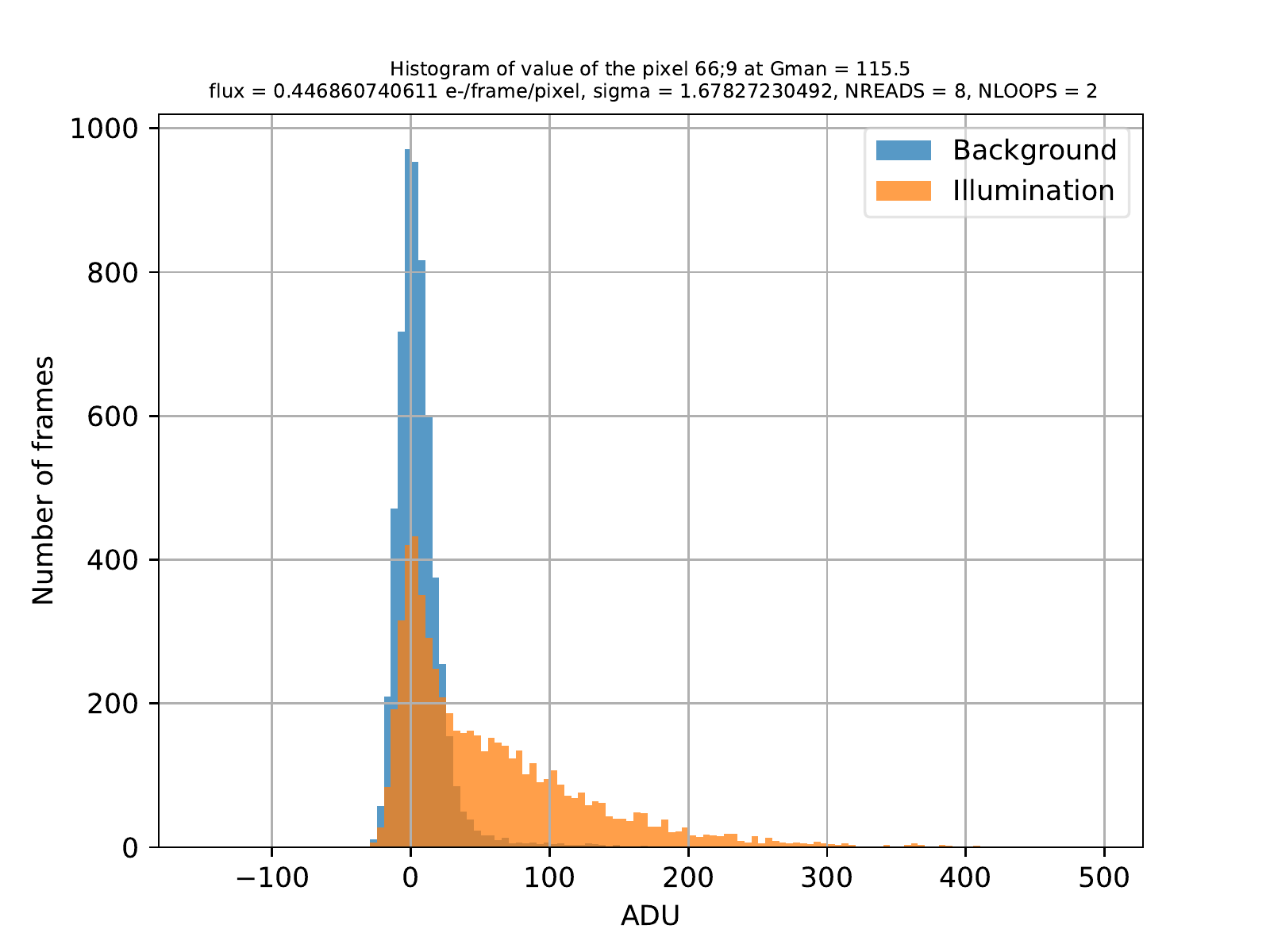}
\caption{Histogram of a temporal sequence}\label{fig:histex}
\end{figure}

\begin{figure}[!h]
\centering
\includegraphics[scale=0.6]{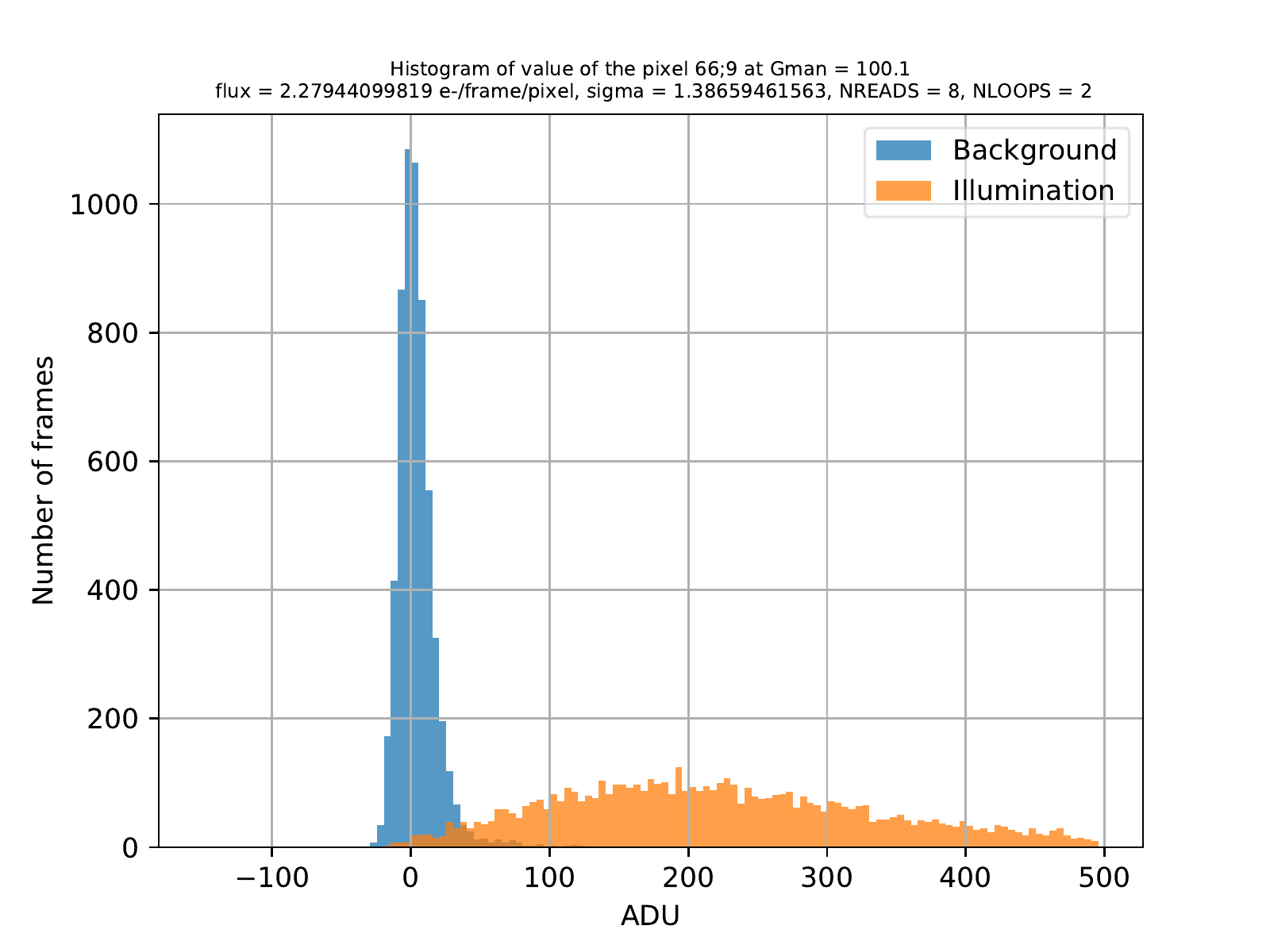}
\caption{Histogram of the remporal sequence shown in Fig.~\ref{fig:dtempseq}}\label{fig:histHF}
\end{figure}

Figure~\ref{fig:histex} is the histogram of the value in the temporal sequence of Fig.~\ref{fig:dtempseq}. Figure~\ref{fig:histHF} is an histogram of a temporal sequence with higher flux level.
On these histograms, the x-axis is the values of the frames in ADU and the y-axis is the number of frames that have these values. The histograms are binned in the x-axis with a bin width of 5 ADU.

The first thing that we notice in Fig.~\ref{fig:histex} and~\ref{fig:histHF} is that we can't separate the different peaks of the different number of photons in a frame. If we had separated peaks, we could straightforwardly determine the number of photons on each frame by the mean of simple thresholds.

In Fig.~\ref{fig:histex}, we see that the illuminated data in orange have a lower peak at 0-photon event than the background data and a larger spread of the histogram toward the positive values. It confirms that we have some flux in the illuminated data.
We also note a break between the peak at 0-photon events and the remaining of the histogram. We could even guess a bump at $\simeq$ 60 ADU. Assuming that this bump is the signature of the 1-photon events, its position (60 ADU) implies that the gain calibrated by the manufacturer is wrong (Gman = 115.5 ADU/$e^{-}$).

A first conclusion is that the histograms of the 0-photon and 1-photon (and $>$ 1-photon) events actually overlap. This blurring is either the result of a wrong estimation of the gain (the true gain being much lower than expected), and/or an effect of the so-called Excess Noise Factor (ENF).

\begin{figure}[!h]
\centering
\includegraphics[scale=0.7]{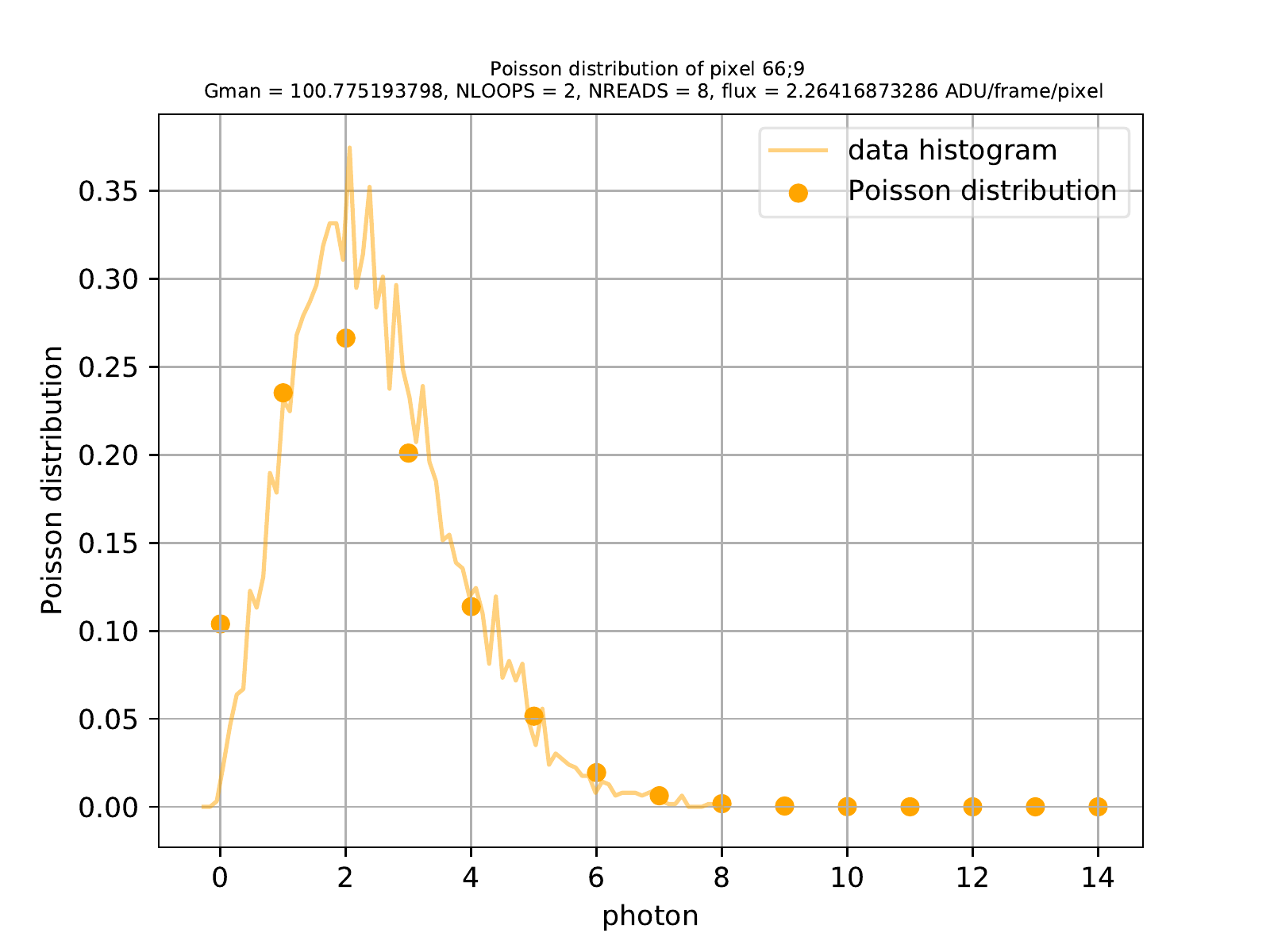}
\caption{Poisson distribution for the flux of the data set of Fig.~\ref{fig:histHF} with expected gain from the manufacturer calibration}\label{fig:photdist2}
\end{figure}

In Fig.~\ref{fig:histHF}, we don't see the 0-photon events peak. It means that we are in a flux regime in which the probability to have 0 photons is negligible. The flux in $e^{-}/frame/pixel$ is given by the mean of the illumination, subtracted from the background, and divided by the gain:
\begin{align}\label{eq:F}
F(e^{-}/frame/pixel) =\frac{\langle ADU_{ill} \rangle - \langle ADU_{BKG} \rangle}{G}  
\end{align}
Assuming the gain calibrated by the manufacturer ($Gman \simeq 100$), we found $F=2.26\,e^{-}/frame/pixel$. In order to compare the observed statistic with the theory, we simulated the Poisson distribution assuming the same flux. Results are shown in Fig.~\ref{fig:photdist2}. The dots are the Poisson distribution. The line is the histogram of the measured data, whose integral has been normalised to one.

At such a low flux-level, the Poisson probability of 0-photon events is still 10$\%$. However, the measured histogram has nearly no 0-photon events. This observation tends to say that the gain calibrated by the manufacturer is wrong, at least in the operational scenario we are using. To our understanding, this is a robust demonstration as it relies entirely on the amount of 0-photon events, which are unaffected by the physics of the amplification. This discrepancy motivated us to derive an independent measurement of the gain, flux and ENF, based on a modeling of the signal distribution.

\subsection{Description of the model}\label{sec:modelpres}

Four parameters can modify the histogram of the data. They are the distribution of counts in the background (e.g histogram of the background), the flux (F), the total gain (G) and the ENF. Note that the information about readout noise is included in the histogram of the background. The gain is a free parameter because the previous section demonstrated that this parameter is false. The flux being unknown, it is a free parameter as well. The ENF is also a free parameter because even if the literature gives us a value between 1.25\cite{CRED-ENF} and 1.3\cite{gert2016}, we would like to confirm it.

The rational of our analysis is that at low flux, the ratio between the 0-photon events and the other events would constrain the Poisson distribution, so the flux. If the flux is constrained, the gain would be constrained as well thanks to eq.~\ref{eq:F}. The ENF would then be constrained by the global shape of the histogram.

The distribution of the data is modeled with the following formula:


\begin{equation}\label{eq:model}
\begin{aligned}
&Histo\,(G, F, ENF, adu) = \\
&\overbrace{\sum_{p}}^{(3)} \bigg[\overbrace{Poisson(p, F)}^{(2)} \;.\;\overbrace{M(ENF, adu/G) * M(ENF, adu/G) * ...}^{(1)\;\;\;\; p - 1\; \mathrm{convolutions}}\bigg] \;\;*\;\; \overbrace{ BKG(adu)}^{(4)}.
\end{aligned} 
\end{equation}

(1) The distribution of a $p$-photon event is represented by $p - 1$ convolutions of the gain distribution $M$. (2) This distribution is weighted by the expected fraction of $p$-photon events, according to the Poisson distribution for the desired flux. (3) All the $p$-photon events distributions are summed. Finally, (4) the result is convolved by the histrogram of the background.

We see that the key point of this model is the gain distribution $M$.

\subsubsection{Gain distribution}\label{sec:Gdist}

As shown in Sec.~\ref{sec:apdtech}, the gain and the ENF are intrinsic properties of the gain distribution M.
A Gaussian distribution~\cite{n-photon} can only model low ENF values ($<$ 1.2). Otherwise the fraction of ``negative amplifications'' becomes significant, which is not physical. A decreasing exponential distribution~\cite{4711091} has a constant ENF of 2.0, which is unsuited to model our expected ENF of 1.25.

The Gamma distribution is classicaly used to model the e-APD and the EMCCD\cite{EMCCDgamma}. It describes the process of multiplication of a single electron inside the pixel. The advantage of the Gamma distribution is that it is an asymmetric function and is defined by 2 positive parameters ($k$ and $\theta$). The former parameter describes the shape of the function. The later changes the scale of the function. These 2 parameters have a direct link with the ENF and the mean gain. For this distribution we have
\begin{equation}\label{eq:meangamma}
 G = \langle M \rangle= k \theta
\end{equation}
and
\begin{equation}\label{eq:vargamma}
Var(M) = k \theta^{2}.
\end{equation}
From eq.~\ref{eq:ENFM}, we get
\begin{equation}\label{eq:ENFgamma}
ENF = 1 + \frac{Var(M)}{E(M^{2})} = 1 + \frac{1}{k}.
\end{equation}
So, from eq.~\ref{eq:meangamma} and ~\ref{eq:ENFgamma}, we have
\begin{equation}\label{eq:k}
k = \frac{1}{ENF-1}
\end{equation}
and
\begin{equation}\label{eq:theta}
\theta = G ( ENF - 1).
\end{equation}

\begin{figure}[!h]
\centering
\includegraphics[scale=0.85]{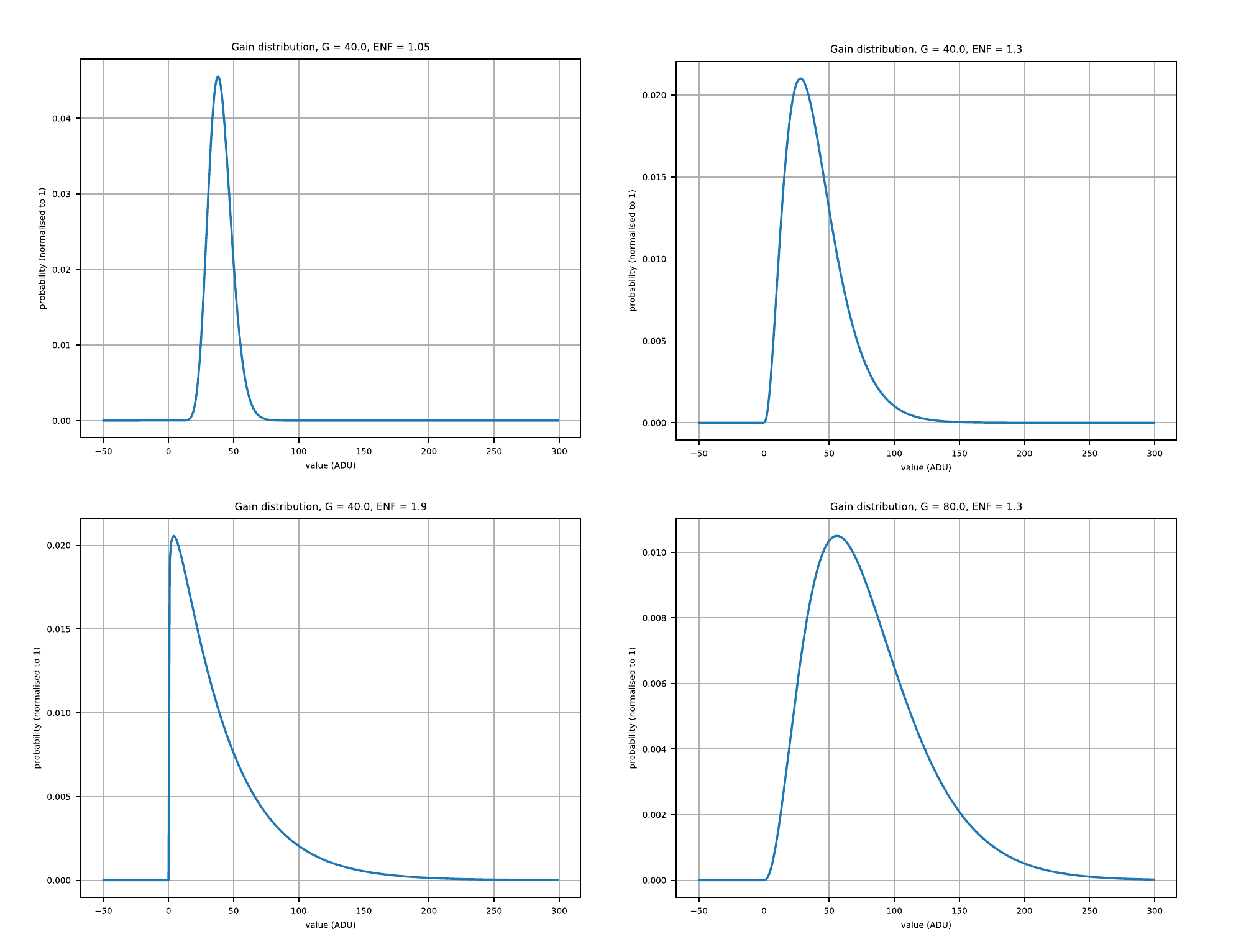}
\caption{Different examples of gain distribution with Gamma distribution}\label{fig:Gdist}
\end{figure}

The Gamma distribution allows a range of ENF value defined by $1.0 < ENF \le 2.0$.
Figure~\ref{fig:Gdist} displays four examples of the gain distribution following a Gamma distribution. On the upper left corner, the ENF = 1.05 and G = 40. On the upper right corner, the ENF = 1.3 and G = 40. On the lower left corner, the ENF = 1.9 and G = 40. On the lower right corner, the ENF = 1.3 and G = 80. Note that the Exponential distribution and the Gaussain distribution are special cases of the Gamma distribution.

As a check, we tried an alternate gain distribution (a Gaussian truncated in 0). The results were relatively similar.

 \newpage

\subsubsection{Minimization strategy}\label{sec:minimsrat}

We use a brute force method, which consists in computing the $chi^{2}$ between the data and the model for a large range of values of the three parameters. The ranges for the flux and the gain are centered around the expected solution. The range for the ENF is the entire range possible by our model of gain distribution (1 to 2).
The result of the brute force exploration is a cube of $chi^{2}$. The minimizer then performs a classical gradient descent toward the best-fit parameters by starting at the position of the minimum of $chi^{2}$ in the cube.

The whole process on the 25 data files takes between 2 and 3 hours with 25 points on the grid for each parameter, and 1 day with 50 points on the grid for each parameter.


\subsection{Results and Interpretations}\label{sec:Resandinter}
\subsubsection{Degeneracy}\label{sec:degen}

\begin{figure}[!h]
\centering
\includegraphics[scale=1]{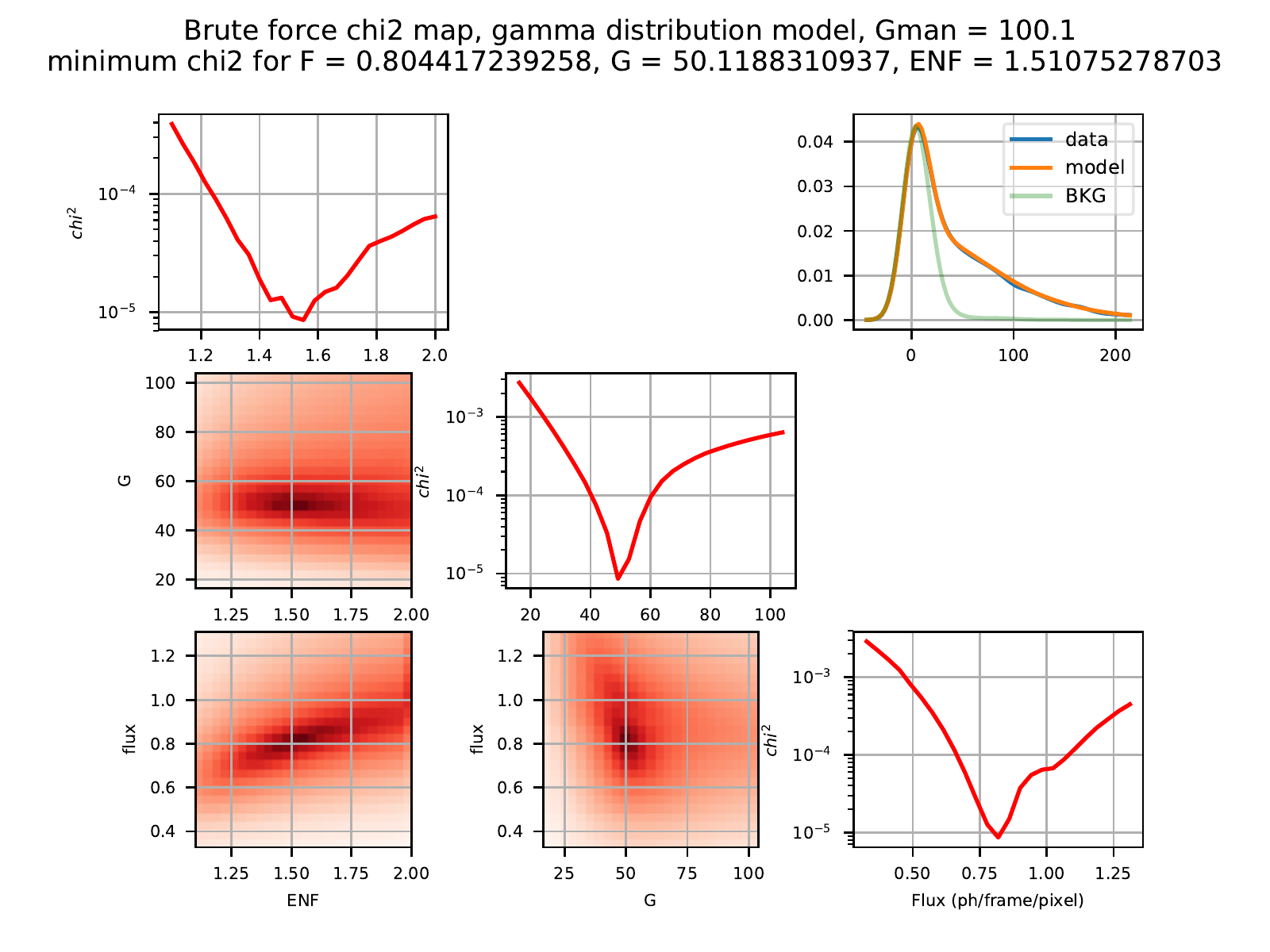}
\caption{$chi^{2}$ minimization result example for low flux}\label{fig:chi2map1}
\end{figure}

Fig.~\ref{fig:chi2map1} shows the results of the minimization for a temporal sequence with a low flux level. In the title, Gman is the expected gain from the manufacturer calibration. F, G and ENF are the flux, the gain and the ENF that minimize the chi2.
The red curves in the diagonal are the minimum $chi^{2}$ for the ENF (top left), the gain (middle) and the flux (bottom right), the two other parameters being free. The maps represent the values of the minimum of $chi^{2}$ as a function of two parameters. The redder it is, the lower is the $chi^{2}$. In the upper right corner, we show the histogram for the illuminated data in blue, the histogram obtained with the parameters that minimize the $chi^{2}$ in orange. The histogram of the background data is shown in light green. This last histogram represents the distribution of the 0-photon events.

The best-fit solution reproduces convincingly the overall shape of the data. Note however that this does not formally demonstrate that our gain model (Gamma distribution) is the only possible one. Interestingly, the solution has no degeneracy, there is only one minimum for each parameter.

\begin{figure}[!h]
\centering
\includegraphics[scale=1]{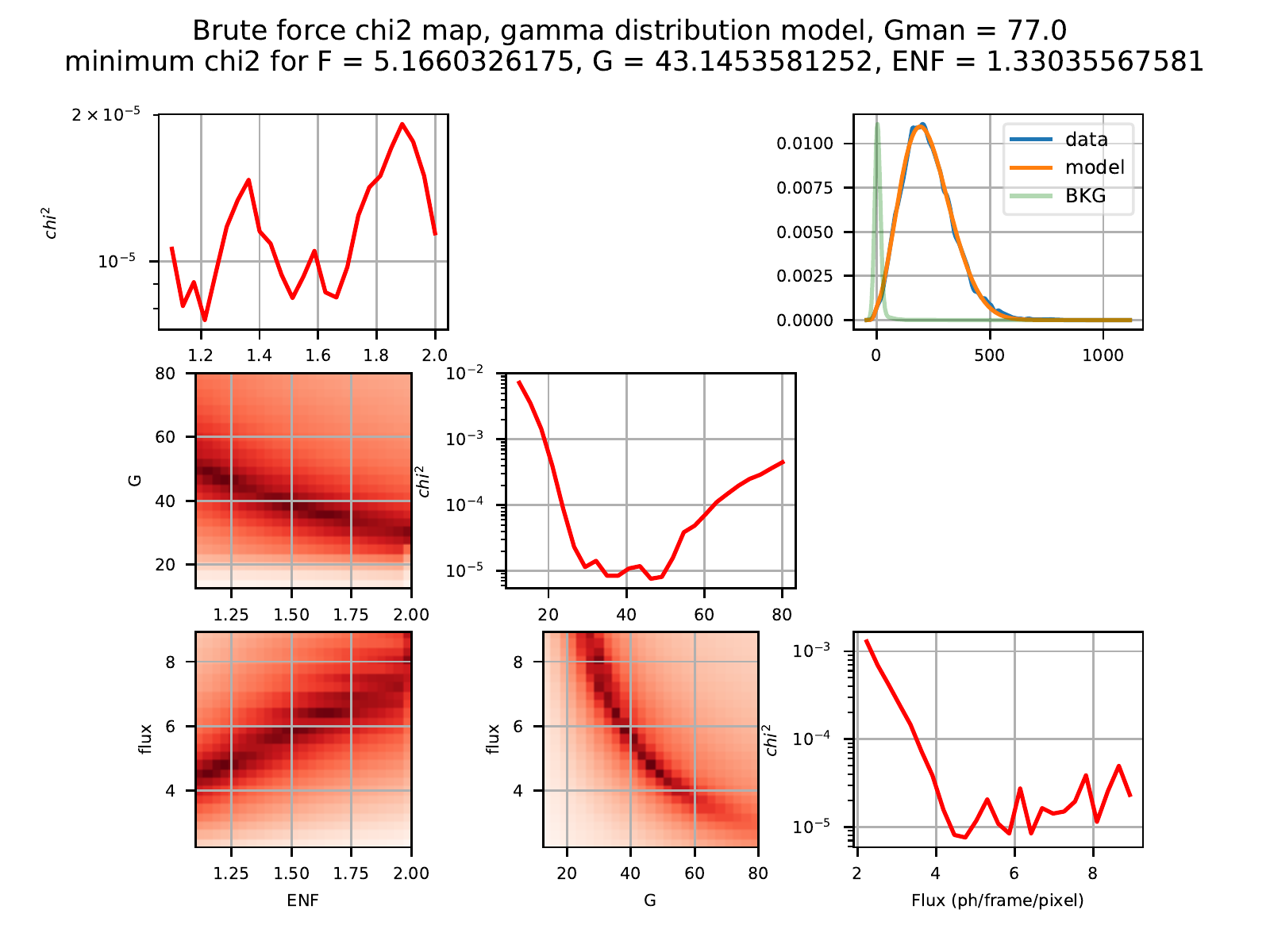}
\caption{$chi^{2}$ minimization result example for high flux}\label{fig:chi2map2}
\end{figure}

Figure~\ref{fig:chi2map2} displays the results for high flux (F $>$ 3 $e^{-}$/frame/pixel) illuminated data. Again, the best-fit solution reproduces convincingly the overall shape of the data. As expected, the solution is more degenerate. The data only constrain a lower limit to the flux, and correspondingly an upper limit on the gain. The excess noise is poorly constrained. The relations between the different parameters follow the expected behavior. The flux is proportional to 1/G in order to reproduce the mean flux in ADU:
\begin{equation}\label{eq:mapFG}
F \,.\, G = \langle ADU \rangle
\end{equation}
And the flux is proportional to the ENF in order to reproduce the total variance:
\begin{equation}\label{eq:mapFENF}
\frac{ENF}{F} = \frac{\mathrm{Var}(ADU)}{\langle ADU \rangle^{2}}.
\end{equation}

\newpage

\subsubsection{Gain}\label{sec:resG}

\begin{figure}[!h]
\centering
\includegraphics[scale = 0.7]{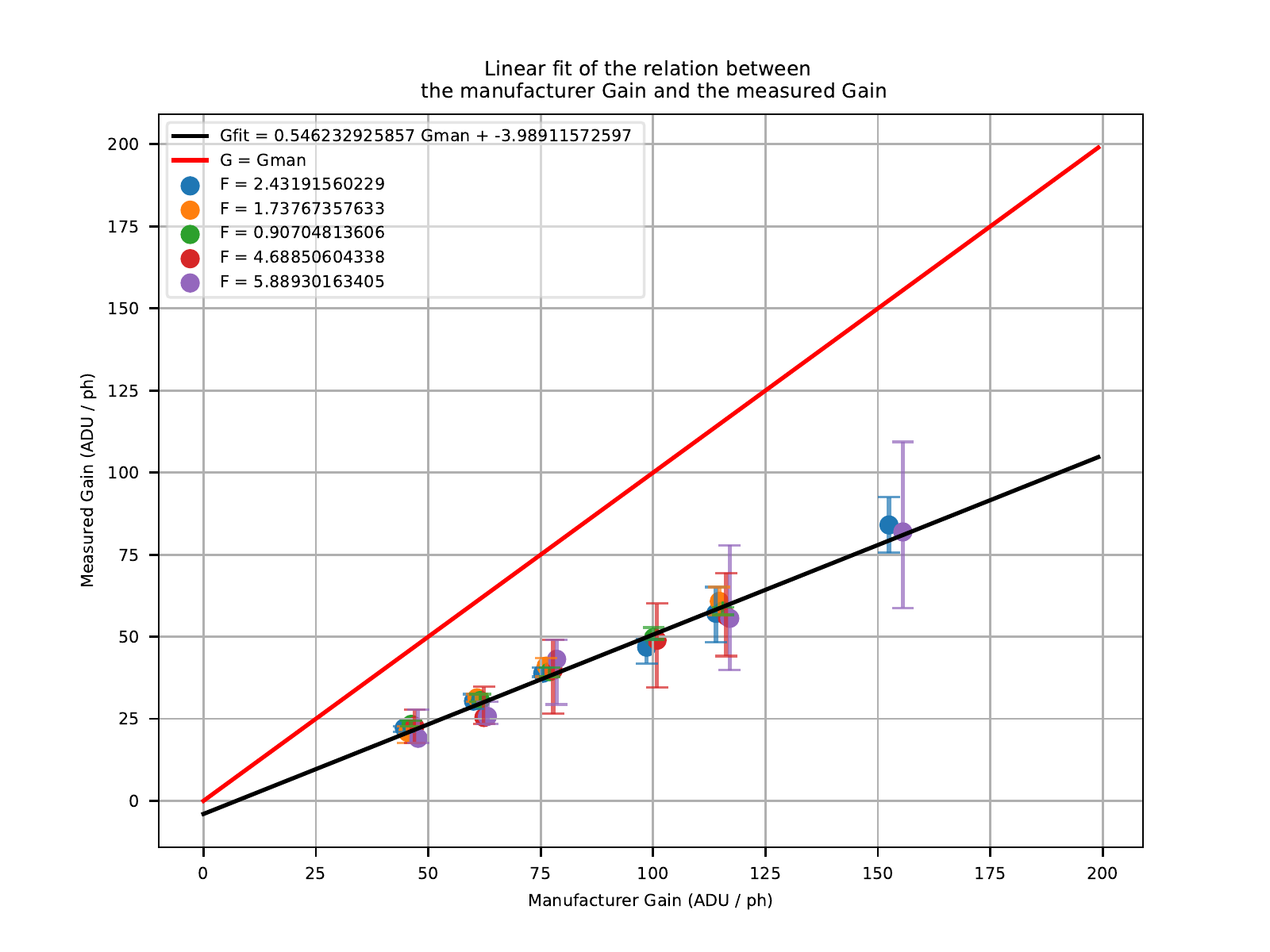}
\caption{Best-fit gain as a function of the expected gain from the manufacturer calibration}\label{fig:GGerrbar}
\end{figure}

Figure~\ref{fig:GGerrbar} shows the best-fit gain versus the expected gain as calibrated by the manufacturer. The black line is a linear fit, while the red line is the expected 1:1 relation. The different colors of the data points show the different flux estimated by the model-fitting.

The results obtained with different flux level are self-consistent.
The linear fit gives us a best-fit gain of 0.55 times the expected gain from the manufacturer calibration. Hereafter, we will call this ratio the gain factor ($Gain\,Factor = G/ Gman$). This value does not depend on the gain nor on the flux. We can notice that the higher the flux level, the larger the error. This is due to the degeneracy that appears at high flux level. Interestingly, the best-fit gains for those high flux data are still consistent with the ones at low flux. This is certainly due to the fact that the minimisation strategy ensures that it always converges toward the global minimum.

\subsubsection{Excess noise factor}\label{sec:resENF}

\begin{figure}[!h]
\centering
\includegraphics[scale = 0.7]{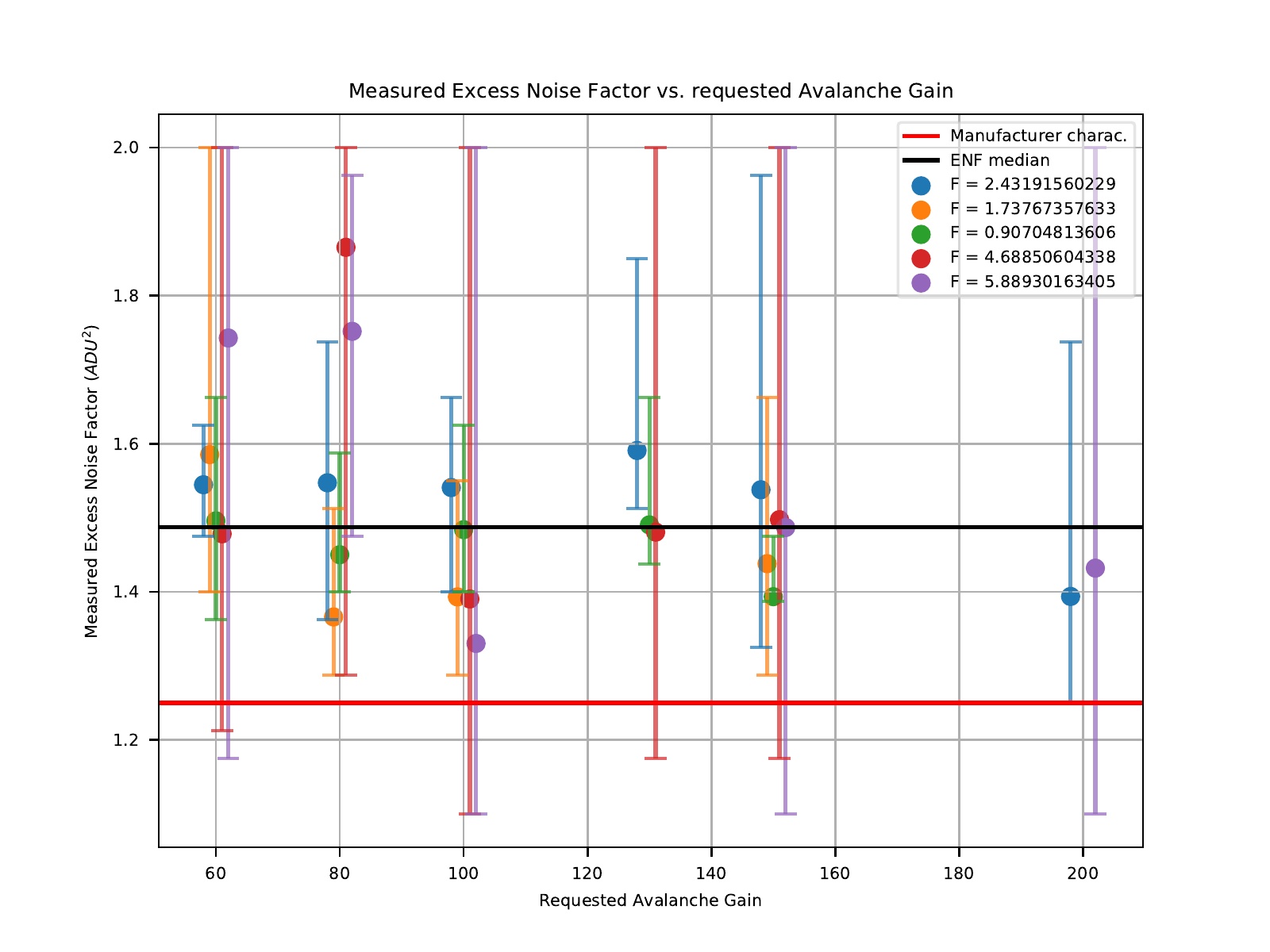}
\caption{Best-fit ENF as a function of the Avalanche Gain}\label{fig:ENFerrbar}
\end{figure}

Figure~\ref{fig:ENFerrbar} displays the measured ENF as a function of the avalanche gain. The different colors of the data points show the different flux estimated by the model-fitting, as in Fig.~\ref{fig:GGerrbar}. The red line shows the value of 1.25 given for the ENF by the manufacturer. The black line shows the median value of the ENF measured by the model.

The measured median value of the ENF is 1.49 $\pm$ 0.12. This is significantly different from the 1.25 expected from the manufacturer specification.
The outliers correspond to observations with high flux, and therefore increased error bars as expected. The error bars all remain compatible with the median ENF = 1.49.

\subsubsection{Flux}\label{sec:resF}

\begin{figure}[!h]
\centering
\includegraphics[scale=0.7]{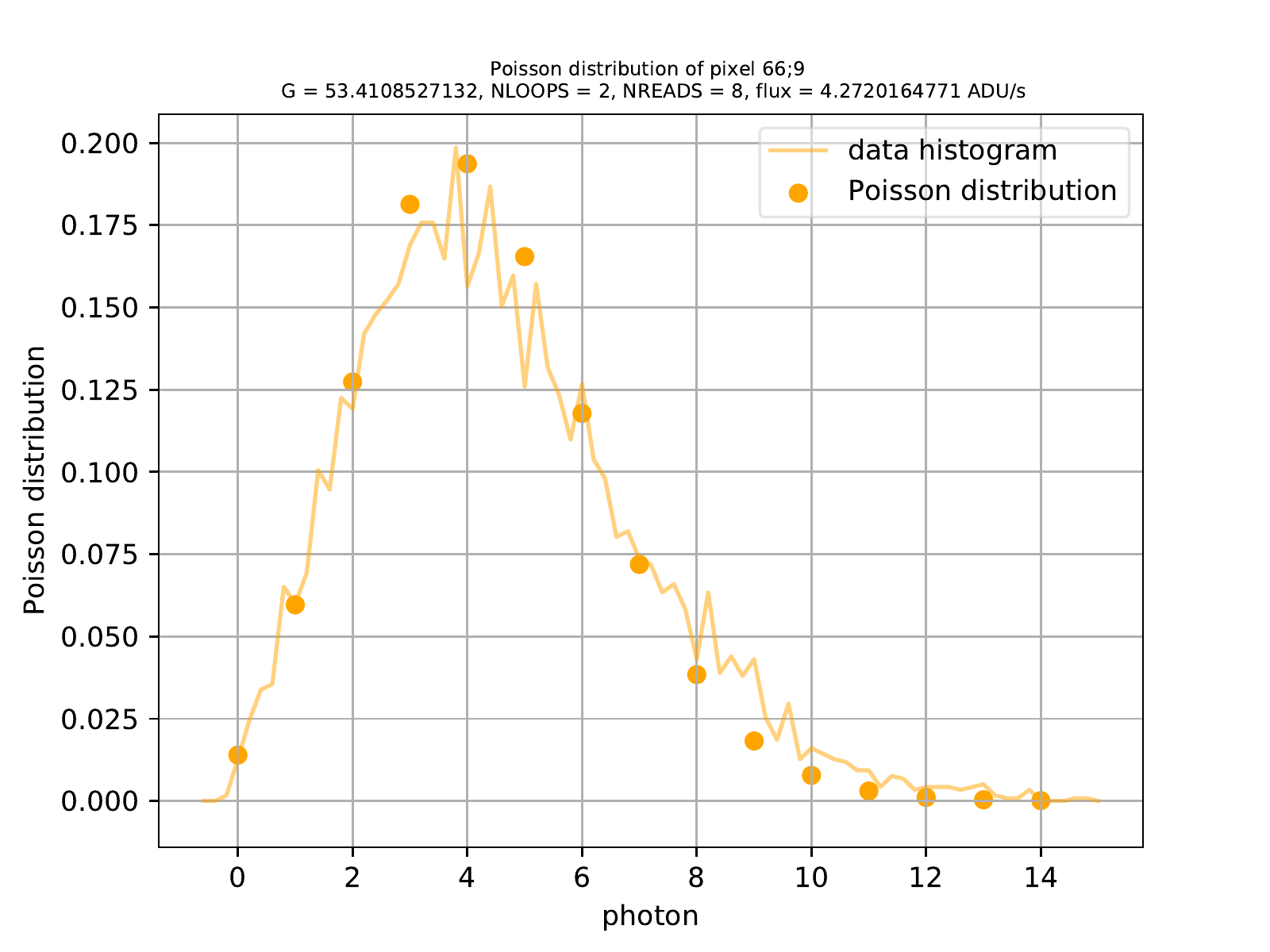}
\caption{Poisson distribution for the flux of Fig.~\ref{fig:histHF} with best-fit gain from the model}\label{fig:poissdist2mes}
\end{figure}

Figure~\ref{fig:photdist2} showed that the Poisson distribution was not consistent with the histogram of the data. We can now make the same comparison, but with the flux computed with the gain from the model that fits the best the data. The histogram used in Fig.~\ref{fig:poissdist2mes} is the same as the one in Fig.~\ref{fig:photdist2}.
We now see that the predicted proportion of 0-photon event is now consistent with the data histogram ($<$0.025 compared to the previous 0.10).

The right part of the histogram is not well reproduced by the Poisson distribution, but this is expected as the latter doesn't include the excess noise factor nor the background. As said previously, the complete model presented in Sec.~\ref{sec:modelpres} reproduces convincingly the observed histograms at all measured flux levels and gains.

\subsubsection{Results for different pixels}\label{sec:ressum}

The results presented above are for the same pixel. Table~\ref{tab:5pixres} summaries the results obtained for 5 different pixels, in order to quickly assess the homogeinity of the result across the detector.

\begin{table}[!h]
\caption{Results of gain and ENF for 5 different pixels} 
\label{tab:5pixres}
\begin{center}       
\begin{tabular}{|l|l|l|} 
\hline
\rule[-1ex]{0pt}{3.5ex}  Pixel coordinate & Gain factor & ENF  \\
\hline
\rule[-1ex]{0pt}{3.5ex} (67; 7) & 0.55 & 1.49   \\
\hline
\rule[-1ex]{0pt}{3.5ex}  (67; 9) & 0.56 & 1.49 \\
\hline 
\rule[-1ex]{0pt}{3.5ex} (66; 9) & 0.55 & 1.49 \\
\hline
\rule[-1ex]{0pt}{3.5ex}  (67; 11) & 0.53 & 1.44  \\
\hline
\rule[-1ex]{0pt}{3.5ex}   (66; 13) & 0.48 & 1.43  \\
\hline\end{tabular}
\end{center}
\end{table}

The coordinates are measured within the 320 x 20 pixel window. The gain factor is computed with the linear fit of the gain of the best-fit model as a function of the manufacturer gain (Fig.~\ref{fig:GGerrbar}). The ENF is obtained with the median of the different ENF measured by the model for various fluxes (Fig.~\ref{fig:ENFerrbar}). The plots shown in previous sections were for pixel (66; 9). The table~\ref{tab:5pixres} shows that we obtain similar values for other pixels. The mean value for the gain factor on these pixels is 0.53 $\pm$ 0.04. The mean value of the ENF of the detector for these pixels is 1.47$\pm$ 0.03.

\section{C-RED ONE OPTIMIZATION}\label{sec:CREDcharac}

\subsection{Choice of the readout mode}\label{sec:romode}

For the most common observational mode, the on-sky observations make use of 320 x 60 pixels. The readout mode chosen is the IOTA mode with Nreads = 8 and Nloops = 6. We found that averaging more than 8 samples doesn't improve the readout noise significantly. Nevertheless, we take more than 8 samples in order to lower the maximum frame rate to $\simeq$ 300 Hz. This frame rate is still faster than the typical bad seeing coherence time ($\simeq$ 5 ms). And it generates lower data volume than higher maximum frame rates. In practice, the observations are all made in a non-destructive mode with this maximum frame rate of $\simeq$ 300 Hz, and the effective integration time is defined a-posteriori during the data analysis by the pipeline.

We perform a study of the the readout noise, the dark current and the background level. This study has for purpose to optimize the use of the camera on-sky. To do so, these characteristics have to be measured for different gains and different effective frame rates (e.g the frame rate ultimately used by the pipeline). The goal of this study is to find the best configurations to use for different target magnitudes and seeing conditions.

The data used in this section has been taken with the actual setup for on-sky observations. We used 40 non-destructive ramps of 100 frames each. We recorded data for 16 different avalanche gains from 1 to 175. To get rid of the electronic signal, we filtered the data using the time fourier transform, and filtering the 90 Hz signal. 

\subsection{Dark Current}\label{sec:BKG}

The dark current includes the thermal background and the internal dark of the camera. Its value is computed looking at a room temperature scene ($\simeq 300 K$). We compute the difference between two frames separated by the desired integration time. Then the mean over the non-destructive ramps of the data and the median over the pixels is computed. The different effective frame rate values are obtained by computing the difference for different frame intervals in the same ramp. For example, for the fastest effective frame rate (284 FPS) the difference is computed between the first and the second frame in a same non-destructive ramp. For the slowest effective frame rate (2.9 FPS) the difference is performed between the first and the last frame in the non-destructive ramp. 

\begin{figure}[!h]
\centering
\includegraphics[scale = 0.77]{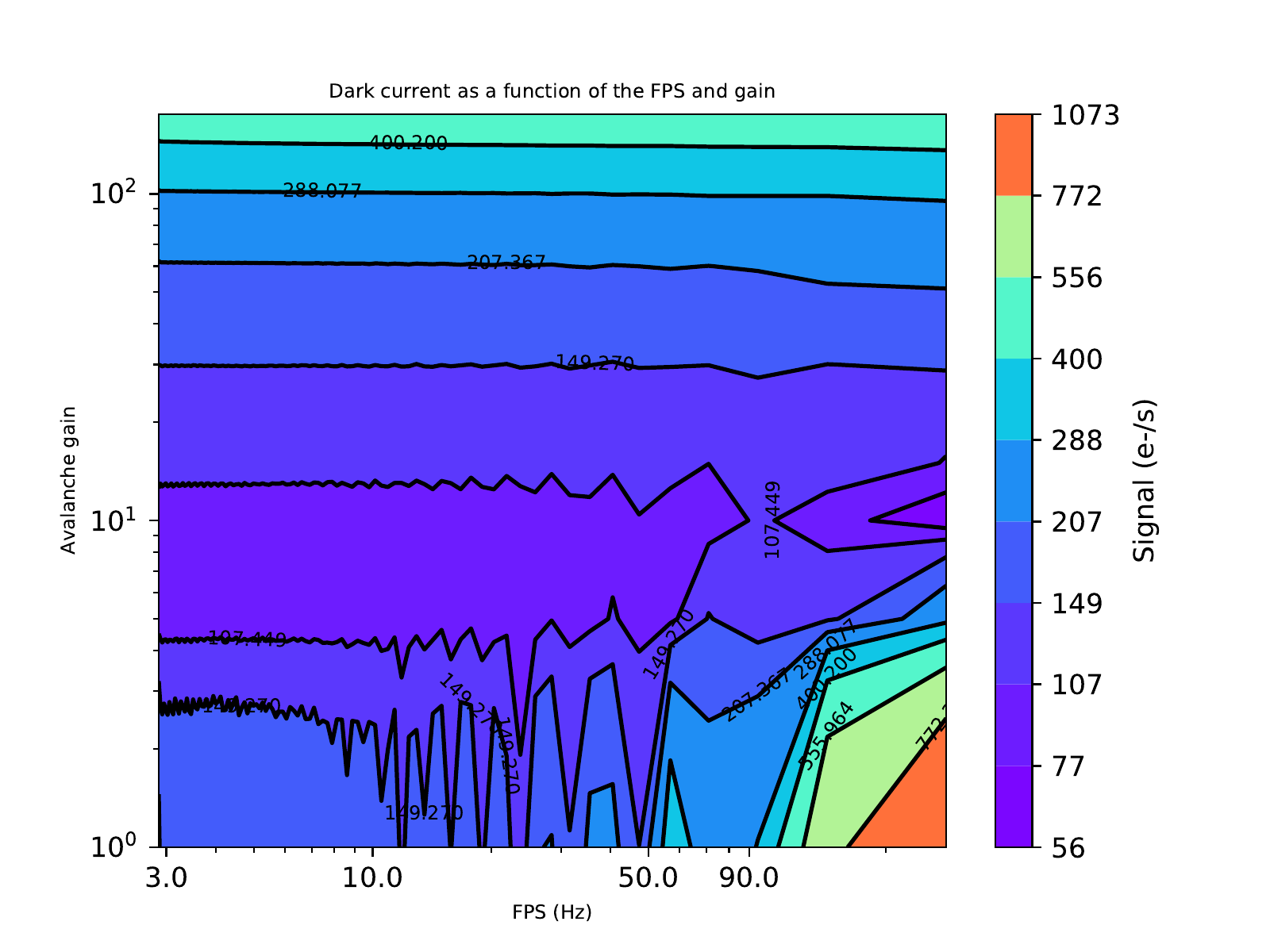}
\caption{Dark current as a function of the avalanche gain and the effective frame rate}\label{fig:contourBKG}
\end{figure}

In Fig.~\ref{fig:contourBKG}, the x-axis is the effective frame rate in FPS and the y-axis is the avalanche gain. Both axes are in logarithmic scale. The different values for the dark current are shown with the logarithmic color scale in $e^{-}$/s. At low gain and high frame rate, the residual effect of the electronic signal is still visible. Therefore, we discard this part of the parameter space from the analysis.

\begin{figure}[!h]
\centering
\includegraphics[scale = 0.7]{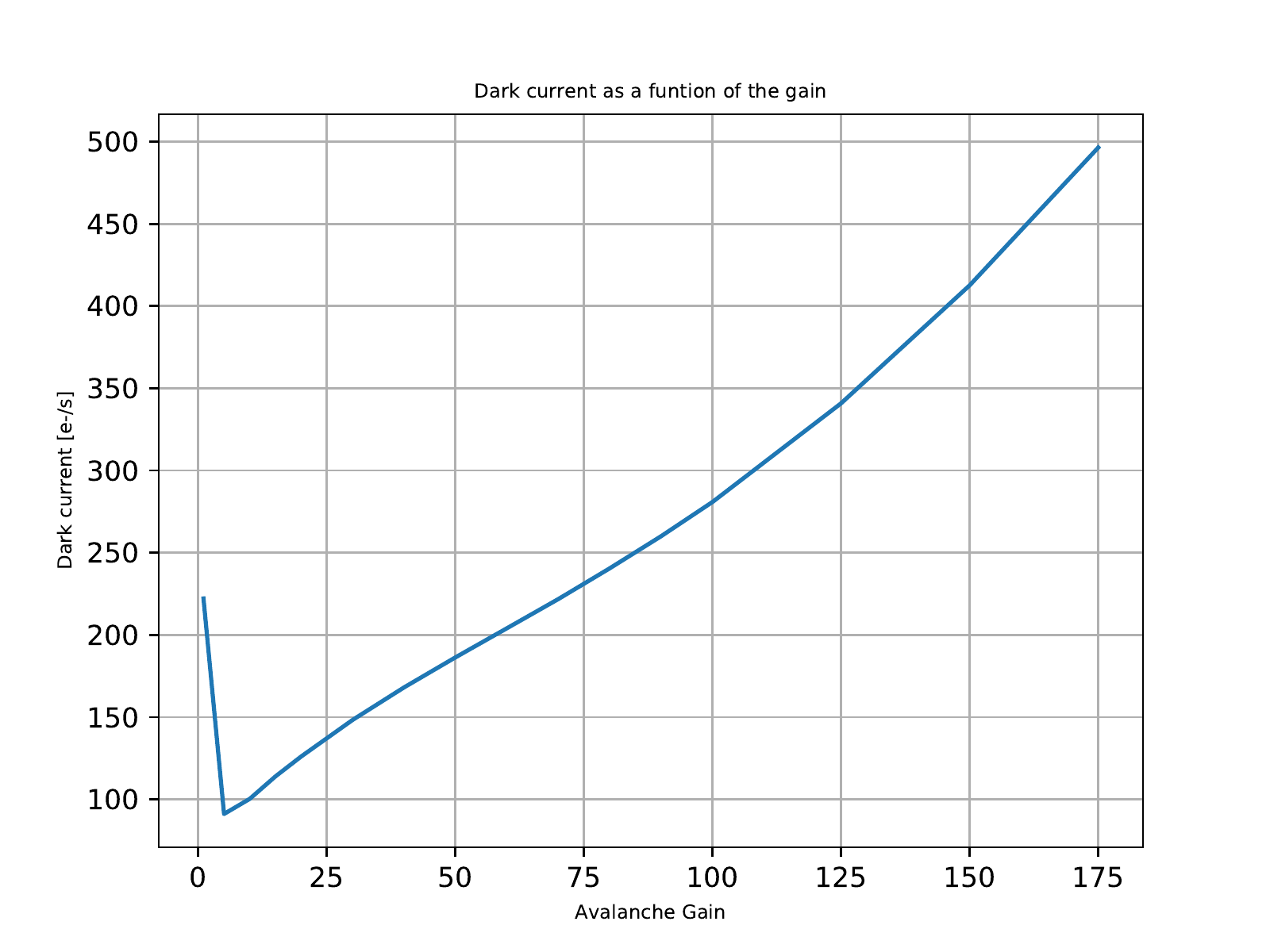}
\caption{Dark current as a function of the avalanche gain and the effective frame rate}\label{fig:BKGvsG}
\end{figure}

As expected, the dark current is flat for the same gain and at different effective frame rate. Figure~\ref{fig:BKGvsG} shows a cut of Fig.~\ref{fig:contourBKG} at the lowest FPS (2.9 FPS). The x-axis is the avalanche gain and the y-axis is the dark current in $e^{-}$/s. The minimum dark current is 90 $e^{-}$/s. It is consistent with the theoretical calculation ($\simeq$ 100 $e^{-}$/s). This calculation takes into account the room temperature, the quantum efficiency of the detector, the transmission of the different filters of the camera and the f-number of the camera (f/4). The 10\% difference can be due to the fact that the total transmission of the camera might be lower than expected. A measure of the transmission of the MYSTIC camera has been performed. It shows that the actual transmission is $\simeq$10$\%$ lower than the expected value. The same measure still has to be done on the MIRC-X camera to confirm that effect. 

The minimum of dark current is obtained for an avalanche gain of 5 and increases for larger gain. This increase is certainly due to tunnel effect\cite{2018arXiv180508419G} because of the high bias voltage for those avalanche gains.

\subsection{Total Noise}\label{sec:ttnoise}

\begin{figure}[!h]
\centering
\includegraphics[scale = 0.5]{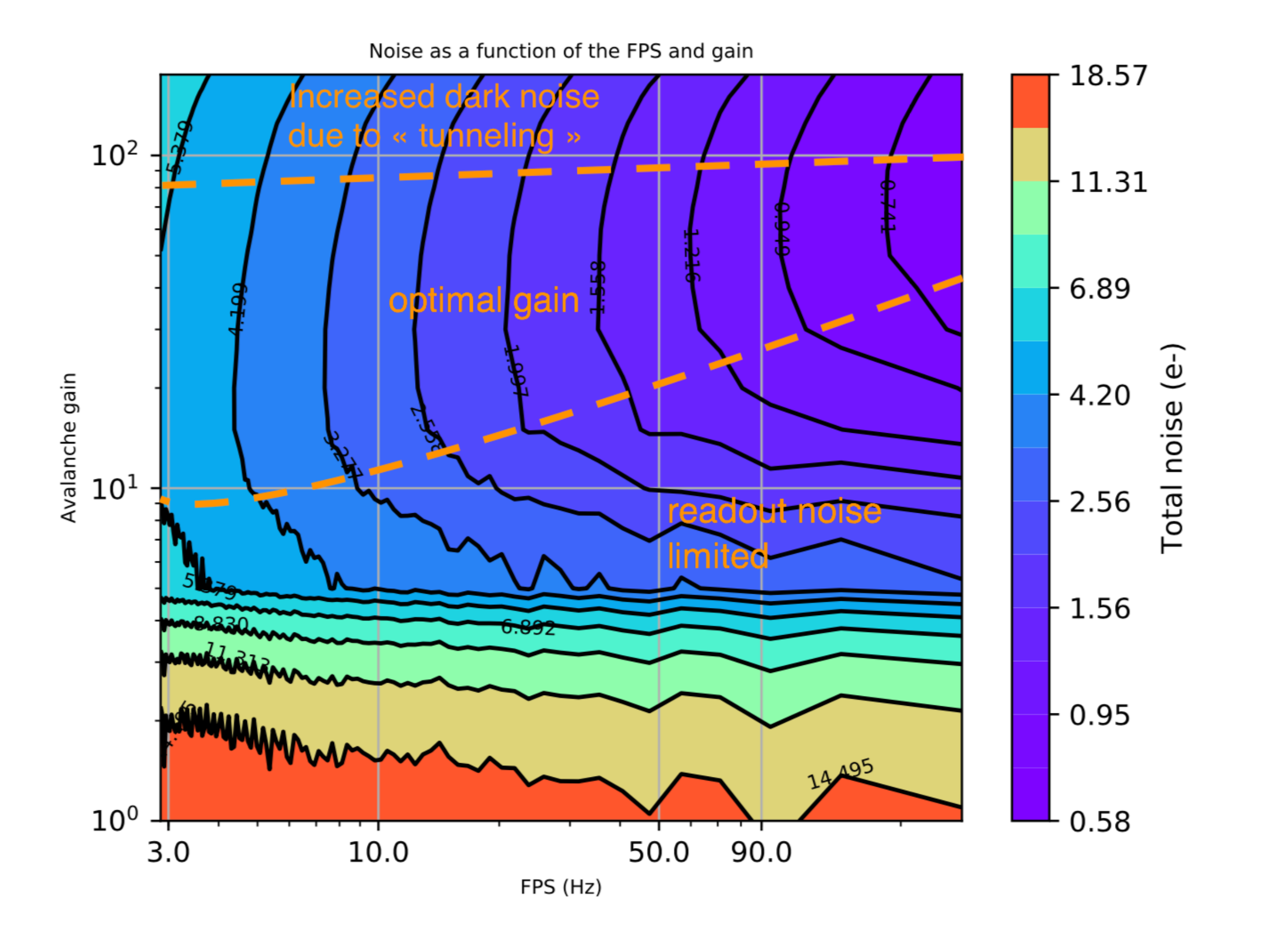}
\caption{Total noise as a function of the avalanche gain and the effective frame rate}\label{fig:contourRMS}
\end{figure}

We compute the total noise as a function of the effective frame rate and the gain following the same method than for Dark current. We performed the standard deviation over the 40 non-destructive ramps of the data for each pixel, and we took the median value of this standard deviation over the pixels. Results are presented in Fig.~\ref{fig:contourRMS}. The x-axis is the frame rate in FPS and the y-axis is the avalanche gain. Both axes are in log scale. The different values for the noise are shown with the logarithmic color scale in $e^{-}$/s. The minimum and maximum number on the color scale are the actual minimum and maximum values of the noise. 

We can reach a sub-electron noise for the highest frame rates and for gain between avalanche gains of 20 and 175. For gain 5 and below, the noise doesn't change with the frame rate. It means that we are limited by the readout noise. Between gain 5 and 10, the noise starts to change with the frame rate, especially for frame rates $<$ 10 FPS. It means that we start to be limited by the background noise, but not totally. At those gains, we are still limited by the camera performances for frame rates $>$ 50 FPS (area called "readout noise limited" in Fig.~\ref{fig:contourRMS}). Between gains 10 and 100, the higher the gain is, the higher is the frame rate for which we are fully background limited. This limit is shown by the bottom orange dashed lines. From the gain 60 to the gain 100, we are still background limited at the maximum frame rate used (284 FPS). We see that at avalanche gain higher than 100, the noise start to increase. The top orange dashed line shows this limit as a function of the frame rate. This is due to the fact that the dark current increases actually more for avalanche gains higher than 100.

The area between the two dashed lines shows the ranges of optimal gains as a function of the frame rate. Bright objects can make use of the lowest integration time ($\simeq 300$ FPS). The optimal gains are between 60 and 100. The optimal gain is 60. Faint objects typically require to integrate for 100\,ms. At this frame rate of 10 FPS, the optimal gain is 30. However, the difference between gain 30 and gain 60 is negligible. Altogether, we recommend to always use a gain of 60 for the observations, regardless of the final integration time that may be used in the post-processing. This conclusion is very convenient from the operational point of view.

\subsection{Apparent sub-Poisson statistic}

The noise at the lowest frame rate and with the optimal gain is 5.6 $e^{-}$ RMS. It corresponds to a flux of $5.6^{2} \times{} 2.9 \simeq 91$ $e^{-}$/s. This result is the same than the minimum dark current of Fig.\ref{fig:BKGvsG}, considered to be purely the external background. It confirms that the background dominates the noise at the optimal gains.

However, the noise at gain higher than 70 is lower than the Poisson noise expected for the corresponding dark current. For instance, at the gain 80 and 2.9 FPS, we measure $\simeq$ 260 $e^{-}$/s, so $\simeq$ 90 $e^{-}$. The expected Poisson noise is then $\sqrt{90} \simeq 9.5$ $e^{-}$. However we measure a noise of 5.6 $e^{-}$. This result could be due to the fact that part of the signal might be less amplified. Indeed, as explained in sec.~\ref{sec:BKG}, some signal can come from electrons generated by tunneling effect, inside the multiplication region of the pixel. Those electrons can be generated from any part of this region including at the bottom. The signal of those electrons would not be, or partially only, amplified by the avalanche process. So, the conversion from ADU to $e^{-}$, involving the gain, would create an apparent sub-Poisson noise.

\subsection{Comparison Between MIRC and MIRC-X}

\begin{figure}[!h]
\centering
\includegraphics[scale=0.8]{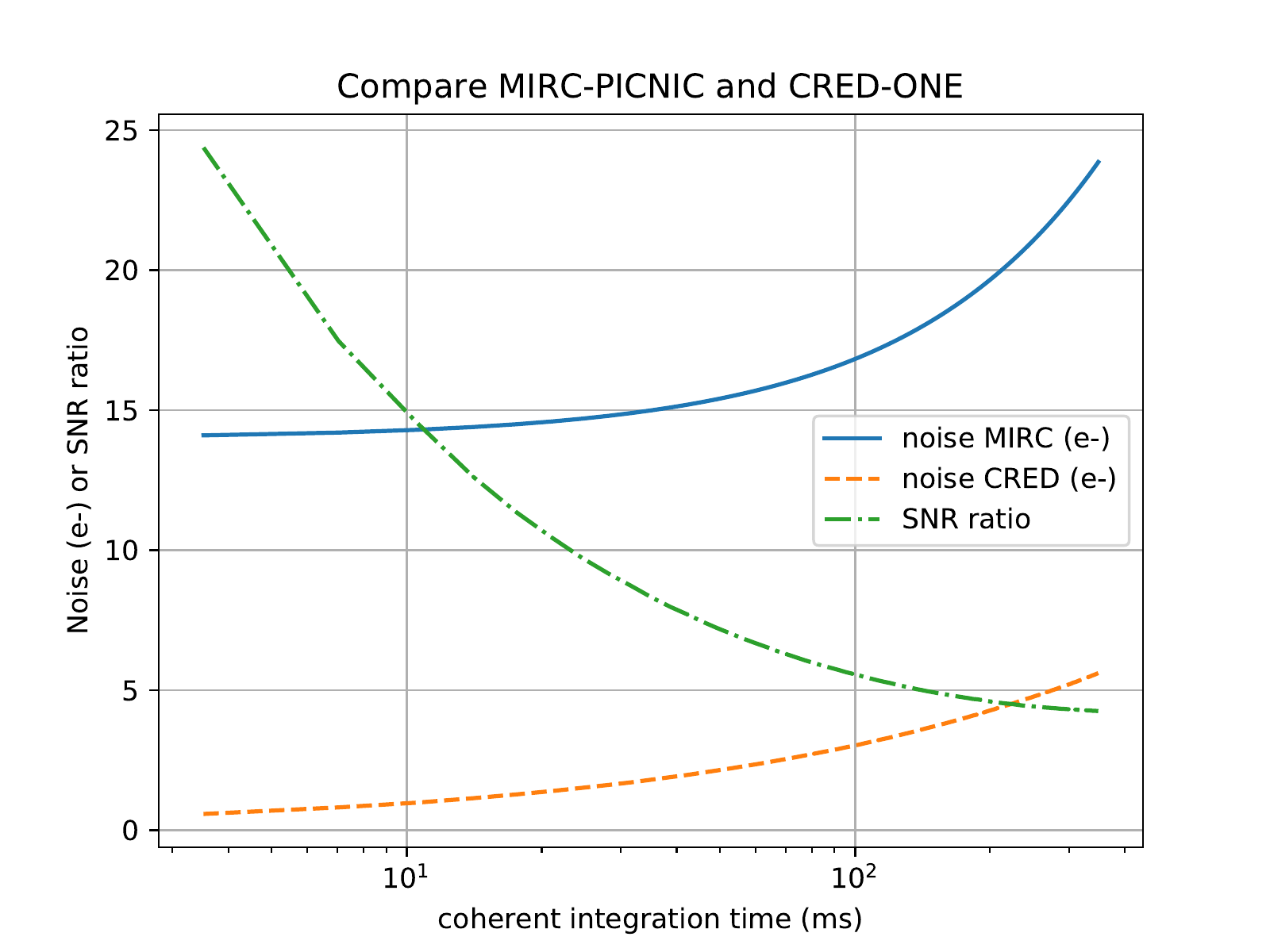}
\caption{Comparision of the PICNIC and the C-RED ONE camera noise}\label{fig:SNRsimu}
\end{figure}

The C-RED ONE camera replaces the previous PICNIC camera of MIRC\cite{PICNIC}. The PICNIC camera had a readout noise of 14 $e^{-}$/read/pixel in operation mode (1 frame is the mean of 4 reads at 1MHz, there is an integration of 5 ms between each frame, and the result is the CDS between frames) and a background of 320 e-/s/pixel. Howerver, the effective noise is a combination of the background noise and the 1/f noise. This total noise can be approximate as a noise coming from a background of 800 $e^{-}$/s/pixel. The C-RED ONE camera readout noise is $<$ 1 $e^{-}$/read/pixel in operation mode (section~\ref{sec:romode}), and its background is 90 $e^{-}$/s/pixel. We expect a better signal to noise ratio (SNR) with a factor of 15 for 10 ms exposure and a factor of 5.5 for 100 ms exposure, as shown in Fig.~\ref{fig:SNRsimu}.

Figure~\ref{fig:SNRsimu} displays the computed noise for the PICNIC camera of MIRC in full line. The measured C-RED ONE noise of MIRC-X for gain 60 is shown in dashed line. The SNR ratio is shown in dashed-dotted line.

The improvement in the SNR is demonstrated by the on-sky performances. The previous MIRC instrument had a magnitude limit of 6.0 in the H-band. Withe the e-APD based detector, the MIRC-X instrument already observed a target of H = 6.5.

\section{CONCLUSIONS}\label{sec:conclusion}

First, the proposed simple model for the camera amplification is able to reproduce adequately the observed distribution of photon events, over a large range of flux and gain. When compared to low-flux data, this model allows to constrain the total gain and Excess Noise Factor. The advantage of this method is to provide a measurement of the total gain in the operational conditions, in term of typical flux and readout mode. Moreover, it does not rely on any preexisting calibration such as the system gain or the ENF.

This study shows that our CRED-ONE camera has a total gain of half the expected gain from the manufacturer calibration. Moreover, the measured ENF (1.47) is higher than the one expected (1.25). As such, photon counting with separated peaks is not possible with the actual performance of our camera.

Compared to the previous PICNIC camera, the CRED-ONE provides an order-of-magnitude improvement in noise performance at short integration times ($\le$ 5 ms). This improvement is reduced to a factor 5 at longer integration times ($\simeq$ 100ms) due to the still relatively high thermal background level. To further improve the performances of the MIRC-X instrument, we have to reduce this background. For that, we would need to implement the camera and the optics in a cryostat, as it is planned for MYSTIC.

\acknowledgments  
We would like to thanks Marc-Antoine MARTINOD and Johan ROTHMAN for their interest in this work and the fruitful discussions that lead to the choice of the Gamma distribution for the gain distribution. We thank also FLI for their collaboration in the characterization and the comprehension of the camera. We also thank the university of Michichan to welcome us and to make possible this collaboration work. We finally thank CHARA and all the people that allows us to take data and observe.
This work has been partially supported by the LabEx FOCUS ANR-11-LABX-0013.
The research leading to these results has received funding from the European Union’s Horizon 2020 research and innovation programme under Grant Agreement 730890 (OPTICON) and Grant Agreement No. 639889 (ERC Starting Grant "ImagePlanetFormDiscs").
This work has been supported by a grant from Labex OSUG@2020 (Investissements d'avenir – ANR10 LABX56).
This work was supported by the Programme National de Physique Stellaire (PNPS) of CNRS/INSU co-funded by CEA and CNES.

\bibliographystyle{spiebib} 
\bibliography{biblio} 

\end{document}